\documentstyle[12pt,twoside]{article}

\def\greaterthansquiggle{\raise.3ex\hbox{$>$\kern-.75em\lower1ex\hbox{$\sim$}}}
\def\lessthansquiggle{\raise.3ex\hbox{$<$\kern-.75em\lower1ex\hbox{$\sim$}}}
\newcommand{\beq}{\begin{equation}}
\newcommand{\eeq}{\end{equation}}
\newcommand{\beqa}{\begin{eqnarray}}
\newcommand{\eeqa}{\end{eqnarray}}
\newcommand{\beqan}{\begin{eqnarray*}}
\newcommand{\eeqan}{\end{eqnarray*}}
\newcommand{\ba}{\begin{array}}
\newcommand{\ea}{\end{array}}

\def\nz{\ifmmode {I\hskip -3pt N} \else {\hbox {$I\hskip -3pt N$}}\fi}
\def\zz{\ifmmode {Z\hskip -4.8pt Z} \else
       {\hbox {$Z\hskip -4.8pt Z$}}\fi}
\def\qz{\ifmmode {Q\hskip -5.0pt\vrule height6.0pt depth 0pt
       \hskip 6pt} \else {\hbox
       {$Q\hskip -5.0pt\vrule height6.0pt depth 0pt\hskip 6pt$}}\fi}
\def\rz{\ifmmode {I\hskip -3pt R} \else {\hbox {$I\hskip -3pt R$}}\fi}
\def\cz{\ifmmode {C\hskip -4.8pt\vrule height5.8pt\hskip 6.3pt} \else
       {\hbox {$C\hskip -4.8pt\vrule height5.8pt\hskip 6.3pt$}}\fi}
\newtheorem{theorem}{Theorem}
\newtheorem{definition}{Definition}
\newtheorem{lemma}{Lemma}

\normalsize
\begin{document}

\begin{titlepage}
\begin{center}
{\Large \bf On energy-momentum
spectrum of stationary states with 
nonvanishing current 
on 1-d lattice systems 
}\\[24pt]
Takayuki Miyadera \\
Department of Information Sciences \\
Tokyo University of Science \\
Noda City, Chiba 278-8510,
Japan
\vfill
{\bf Abstract} \\
\end{center}
On one-dimensional two-way infinite quantum lattice system,
a property of translationally invariant stationary states with nonvanishing 
current expectation is investigated.
We consider GNS representation with respect to such a state,
on which we have a group of space-time translation unitary operators.
We show 
that spectrum of the unitary operators, energy-momentum
spectrum with respect to the state, has a singularity at 
the origin.
\vfill
\small
e-mail: miyadera@is.noda.tus.ac.jp
\end{titlepage}

\section{Introduction}

Recently a lot of researchers get interested in nonequilibrium
states \cite{HoA,Ta,Shmz,Oji,Fr,JP}. 
In spite of their efforts, few things have been known rigorously.
The situation is contrast to equilibrium state. In equilibrium state
business, for instance, properties of energy momentum spectrum have been
well understood \cite{Haag,BR}. We, in the present paper, study
rigorously a property of energy-momentum spectrum with respect to
nonequilibrium steady states. To put it concretely, we consider a
one-dimensional lattice system with nearest neighbor interaction, and
translationally invariant stationary 
states with nonvanishing current on it.
Because of the space-time translational invariance of the state, on 
its GNS
representation there exists a group of unitary operators whose spectrum is
called energy-momentum spectrum. We show that the spectrum has singularity
at the origin thanks to 
the nonvanishing nature of the current. Our discussion
is model independent and general. 

The paper is organized as follows. In the next section we
briefly introduce one-dimensional lattice system and define a state with
nonvanishing current what we are interested in. In section \ref{sect:em}%
, we show our main theorem. 
\section{States with nonvanishing current on 1-d lattice systems}
We deal with a one-dimensional two-way infinite quantum 
lattice system. 
To each site $x\in {\bf Z}$ a Hilbert space ${\cal H}_{x}$ which is
isomorphic to ${\bf C}^{N+1}$ is attached and observable algebra at site 
$x$ is a matrix algebra on ${\cal H}_{x}$ 
which is denoted by ${\cal A}(\{x\})$. 
The observable algebra on a finite set $\Lambda \subset {\bf Z}$ is a matrix
algebra 
on $\otimes _{x\in \Lambda }{\cal H}_{x}$ and denoted by 
${\cal A}(\Lambda )$. 
Natural identification can be used to derive an inclusion property 
${\cal A}(\Lambda _{1})\subset {\cal A}(\Lambda _{2})$ for 
$\Lambda _{1}\subset\Lambda _{2}$. 
The total observable algebra is a norm completion of sum
of the finite region observable algebra, 
${\cal A}:=\overline{\cup _{\Lambda
:\mbox{\small{finite}}}{\cal A}(\Lambda )}^{\Vert \Vert }$, which becomes a $%
C^{\ast }$ algebra. (For detail, see \cite{BR}.) 

To discuss the dynamics, we need $\{\alpha_t\}_{t\in {\bf R}}$,
a one-parameter $\ast $%
-automorphism group on ${\cal A}$, which we assume is induced by a local
interaction. In the present paper, for simplicity, we assume that 
the interaction
is nearest neighbor one. 
That is, for each $x\in {\bf Z}$ there exists a self adjoint element $%
h_{x,x+1}\in {\cal A}([x,x+1])$, and the local Hamiltonian with respect to
each finite region $\Lambda $ is defined by 
\begin{eqnarray*}
H_{\Lambda }:=\sum_{\{x,x+1\}\subset \Lambda }h_{x,x+1}.
\end{eqnarray*}
Moreover, we assume translational invariance of the interaction.
That is, 
\begin{eqnarray*}
\tau _{x}(h_{y,y+1})=h_{x+y,x+y+1},
\end{eqnarray*}
holds for each $x,y\in {\bf Z}$ where $\tau _{x}$ is a space translation $%
\ast $-automorphism. \newline
The Hamiltonian defines a one-parameter $\ast $-automorphism $\alpha _{t}$
by 
\begin{eqnarray*}
\frac{d\alpha_{t}(A)}{dt}:=-i\lim_{\Lambda \to {\bf Z}}
[\alpha_{t}(A),H_{\Lambda }]
\end{eqnarray*}
for each local element $A\in {\cal A}$. Hereafter, for each local element $%
A\in {\cal A}$, we employ the notation $A(t):=\alpha_{t}(A)$.

To define a current operator, we assume existence of local
charge operators. Namely there exists a self-adjoint operator $n_{x}\in 
{\cal A}(\{x\})$ for each $x\in {\bf Z}$ with $\tau_{x}(n_{0})=n_{x}$, and
we put $N_{\Lambda }:=\sum_{x\in \Lambda }n_{x}$ for each finite region $%
\Lambda $. The charge defines a one-parameter $\ast $-automorphism group on
the observable algebra by 
\begin{eqnarray*}
\frac{d\gamma_{\theta }(A)}{d\theta }=i\lim_{\Lambda \to {\bf Z}%
}[N_{\Lambda },\gamma_{\theta }(A)].
\end{eqnarray*}
We assume $N_{\Lambda }$ is conserved with respect to $H_{\Lambda }$, that
is, 
\begin{eqnarray}
[N_{\Lambda },H_{\Lambda }]=0
\label{conserv}
\end{eqnarray}
holds for each finite region $\Lambda $. In particular, putting 
$\Lambda=[x,x+1]$, we obtain a
commutator, 
\begin{eqnarray}
[h_{x,x+1},n_{x}+n_{x+1}]=0.
\label{com}
\end{eqnarray}
On the other hand, by letting $\Lambda \to {\bf Z}$ 
this relation derives a purely
algebraic relation, 
\begin{eqnarray*}
\alpha_{t}\circ \gamma_{\theta }=\gamma_{\theta }\circ \alpha_{t}.
\end{eqnarray*}
With this algebraic relation, $\gamma_{\theta }$ is called a (continuous) 
symmetry
transformation.

On this setting, {\it electric current}  (hereafter we simply
call it as {\it current}) between sites $x$ and $x+1$ is defined by 
\begin{eqnarray*}
j_{x,x+1}:=-i[n_{x+1},h_{x,x+1}]=i[n_{x},h_{x,x+1}],
\end{eqnarray*}
where the second equality is due to (\ref{com}). If we consider the equation
of motion for the charge contained in a finite region $\Lambda:=[-L,0]$, we
obtain 
\begin{eqnarray}
\left. \frac{d \alpha_t(N_{\Lambda})}{dt}\right|_{t=0} =j_{-L-1,-L} -j_{0,1},
\label{cont}
\end{eqnarray}
which corresponds to a continuity equation in continuum case. The following
observation is significant to derive our main theorem. Thanks to (\ref
{conserv}), the current at the origin can be rewritten for any $L$ and $M$
satisfying $L\geq M>0$ as 
\begin{eqnarray*}
j_{0,1}:= i [N_{[-L,0]}, H_{[-M,M+1]}].
\end{eqnarray*}
The above seemingly abstract setting has physically interesting
examples. For instance, interacting fermion system is on the list. 
For each $x\in {\bf Z}$, charge $n_{x}:=c_{x}^{\ast }c_{x}$ and 
$h_{x,x+1}=-t(c_{x+1}^{\ast }c_{x}+c_{x}^{\ast }c_{x+1})-\mu
n_{x}+v(1)n_{x}n_{x+1}$ 
gives a nearest neighbor Hamiltonian. 
The current at the origin is calculated as 
$j_{0,1}=it(c_{1}^{\ast}c_{0}-c_{0}^{\ast }c_{1})$. 
Heisenberg model can be another example. $%
h_{x,x+1}:=S_{x}^{(1)}S_{x+1}^{(1)}+S_{x}^{(2)}S_{x+1}^{(2)}+\lambda
S_{x}^{(3)}S_{x+1}^{(3)}$ and $n_{x}:=S_{x}^{(3)}$ leads the current $%
j_{0,1}=-S_{0}^{(2)}S_{1}^{(1)}+S_{0}^{(1)}S_{1}^{(2)}$. 

 Now we introduce states which we are interested in. 

\begin{definition}
A state $\omega$ over two-way infinite lattice system ${\cal A}$
is called a {\it translationally invariant stationary state with
nonvanishing current} (a state with nonvanishing current, for short) iff the
following conditions are all satisfied: 
\begin{item}
\item[(1)] 
$\omega$ is stationary, i.e., $\omega\circ \alpha_t =\omega$ for all $t$. 
\item[(2)]
$\omega$ is translationally invariant. i.e,. $\omega\circ \tau_x =\omega$
for all $x$.
\item[(3)]
$\omega$ gives non-vanishing expectation of the current,
i.e., 
$
\omega(j_{0,1}) \neq 0.
$
\end{item}
\quad 
\end{definition}

Here we do not impose any other condition, stability for
instance. Our definition hence might allow rather unphysical states which
should be hardly realized. It, however, contains physically interesting
states like nonequilibrium steady states obtained by inhomogeneous initial
conditions which were discussed in \cite{HoA,Ta}. 

We put a GNS representation with respect to a state with
nonvanishing current $\omega $ as $({\cal H},\pi ,\Omega )$. Since we fix a
state $\omega $, indices showing the dependence on $\omega $ will be omitted
hereafter. Moreover we identify $A$ with $\pi (A)$ and will omit to write $%
\pi $. 

Since the state with nonvanishing current $\omega$ is
stationary and translationally invariant, one can define a unitary operator $%
U(x,t)$ for each $x\in {\bf Z},\ t\in {\bf R}$ on ${\cal H}$ by 
\begin{eqnarray*}
U(x,t)A\Omega := \alpha_t\circ\tau_x(A)\Omega
\end{eqnarray*}
for each $A \in {\cal A}$. Thanks to commutativity of time and space
translation, the unitary operators satisfy 
\begin{eqnarray*}
U(x_1,t_1)U(x_2,t_2)=U(x_1+x_2,t_1+t_2)
\end{eqnarray*}
and can be diagonalized into the form:
\begin{eqnarray*}
U(x,t)=\int_{k=-\pi}^{\pi} \int_{\epsilon=-\infty}^{\infty} e^{i(\epsilon
t-kx)}E_{\omega}(dkd\epsilon),
\end{eqnarray*}
where $E_{\omega}(dkd\epsilon)$ is a projection valued measure and called
energy momentum spectrum. 
In the following section, we investigate a property of $%
E_{\omega}(dkd\epsilon)$. %
\section{ Energy momentum spectrum}\label{sect:em} 
In this section we show a singular nature of 
energy-momentum 
spectrum $E_{\omega}(dk d\epsilon)$.
The point of the proof is to estimate the following quantity: 
\begin{eqnarray*}
\int dt(\Omega ,i[N_{[-L,0]},H_{[-M,M+1]}(t)]\Omega )f(t),
\end{eqnarray*}
where $f$ is an arbitrary function with $\mbox{supp}f\subset [ -T,T]$
satisfying $\int dt|f(t)|^{2}<\infty $. 
Although $(\Omega ,i[N_{[-L,0]},H_{[-M,M+1]}(t)]\Omega )$ is not time
invariant, it is almost time invariant for sufficiently large $L$ and $M$. 
To show it, we will employ repeatedly the following lemma. 
\begin{lemma}
\label{th:gp} Let $V(h_{0,1})$ be a quantity which is
determined by the interaction $h_{0,1}$ as 
\begin{eqnarray*}
V(h_{0,1}):=4 (N+1)^{4} e^{2} \Vert h_{0,1}\Vert.
\end{eqnarray*}
For each finite region $\Lambda$, we denote $d(\Lambda):= \mbox{max}%
\{|x-y||\ x,y\in {\Lambda}\}$. Then for all $A\in {\cal A}(\Lambda_1)$ and $%
B\in {\cal A}(\Lambda_2)$ with $0 \in \Lambda_1$ and $0 \in \Lambda_2$ and $x
$ satisfying $|x|-(d(\Lambda_1)+d(\Lambda_2))>0$, an inequality, 
\begin{eqnarray*}
\Vert [\tau_x \alpha_t(A),B]\Vert 
\leq&& 2 (N+1)^{d(\Lambda_1)+d(\Lambda_2)}
\Vert A\Vert \Vert B\Vert d(\Lambda_1)d(\Lambda_2)  \nonumber \\
&& \mbox{exp}\left\{ -|t|(\frac{|x|-(d(\Lambda_1)+d(\Lambda_2))}{|t|} -2
V(h_{0,1}))\right\}
\end{eqnarray*}
holds. 
\end{lemma}

The proof is a direct application of theorem 6.2.11 of \cite{BR},
and is omitted. This lemma guarantees the existence of a {\it finite group
velocity} which is determined by form of the interaction. 
Now we show the following lemma: 
\begin{lemma}\label{th:start} 
For an arbitrary $T>0$ and an arbitrary
function $f$ with the support $[-T,T]$ satisfying $\int dt |f(t)|^2 <\infty$%
, the following equation holds: 
\begin{eqnarray*}
\lim_{M \to \infty} \lim_{L \to \infty} \int dt i[\hat{N}_{[-L,0]},
\hat{H}_{[-M,M+1]}(t)] f(t) =\sqrt{2\pi} j_{0,1} \tilde{f}(0),
\end{eqnarray*}
where 
$\tilde{f}(\epsilon):= \frac{1}{\sqrt{2\pi}} \int dt f(t)e^{i\epsilon t}$
and $\hat{A}:=A-\omega(A)$ for $A\in{\cal A}$, and the limit is taken with
respect to norm topology. 
Note that the order of the limiting procedures can not be 
exchanged.
\end{lemma}

{\bf Proof:} To estimate the equation, let us first consider
the following quantity. 
\begin{eqnarray}
&&
[N_{[-L,0]},H_{[-M,M+1]}(t)]-[N_{[-L,0]},H_{[-M,M+1]}(0)]  
\nonumber \\
&=&\int_{0}^{t}ds[N_{[-L,0]},\alpha _{s}\left(\left. 
\frac{dH_{[-M,M+1]}(u)}{du}%
\right| _{u=0}\right)]  \nonumber \\
&=&-i\int_{0}^{t}ds[N_{[-L,0]},\alpha _{s}\left(
[H_{[-M,M+1]},H_{[-M-1,M+2]}]\right)]
\label{eq1}
\end{eqnarray}
The term $i[H_{[-M,M+1]},H_{[-M-1,M+2]}]$ expresses time derivative of
energy contained in $[-M,M+1]$ and can be decomposed into in-going and
out-going energy current. That is, in a similar manner with electric
current, we define {\it energy current} at a site $x$ by $%
J_{x}:=i[h_{x-1,x},h_{x,x+1}]\in {\cal A}([x-1,x+1])$%
, then the above term is written as 
\begin{eqnarray*}
i[H_{[-M,M+1]},H_{[-M-1,M+2]}]=-J_{-M}+J_{M+1},
\end{eqnarray*}
and
\begin{eqnarray}
(\ref{eq1})=\int_{0}^{t}ds[N_{[-L,0]},-J_{M+1}(s)+J_{-M}(s)]
\label{eq2}
\end{eqnarray}
holds. Now thanks to spacelike commutativity, $[N_{[-L,0]},J_{M+1}]=0$ holds, 
and we obtain also for $J_{-M}$, 
\begin{eqnarray*}
[N_{[-L,0]},J_{-M}] 
&=&-i[N_{[-L,0]},[H_{[-M,M+1]},H_{[-M-1,M+2]}]] \\
&=&i([H_{[-M,M+1]},[H_{[-M-1,M+2]},N_{[-L,0]}]]+
[H_{[-M-1,M+2]},[N_{[-L,0]},H_{[-M,M+1]}]])
\\
&=&[H_{[-M,M+1]},-j_{0,1}]+[H_{[-M-1,M+2]},j_{0,1}]=0,
\end{eqnarray*}
where we used Jacobi identity for commutators. To estimate (\ref{eq2}), we
bound the deviation for finite $s$ by use of lemma\ref{th:gp} as
\begin{eqnarray}
\Vert [ N_{[-L,0]},J_{M+1}(s)]\Vert  
&\leq &\sum_{-L-M\leq z\leq -M-1}
\Vert [ n_{z},J_{0}(s)]\Vert   \nonumber \\
&\leq &
2(N+1)^{4}\Vert n_{0}\Vert \Vert J_{0}\Vert 3
\sum_{-L-M\leq z\leq -M-1}%
\mbox{exp}\{-|s|(\frac{|z|-4}{|s|}-2V(h_{0,1}))\}  \nonumber \\
&\leq &
6(N+1)^{4}\Vert n_{0}\Vert \Vert J_{0}\Vert \frac{e^{-M}}{1-e^{-1}}%
e^{3}e^{2|s|V(h_{0,1})}.  \label{eq21}
\end{eqnarray}
Next we estimate the other term of (\ref{eq2}), 
\begin{eqnarray}
\Vert [ N_{[-L,0]},J_{-M}(s)]\Vert  
&=&\Vert 
[ \alpha_{-s}(N_{[-L,0]}),J_{-M}]\Vert   \nonumber \\
&\leq&\Vert [ N_{[-L,0]},J_{-M}]\Vert +\Vert \int_{0}^{s}du
[\alpha_{-u}\left( \left. \frac{d\alpha _{-t}(N_{[-L,0]})}{dt}\right| _{t=0}
\right),J_{-M}]\Vert   \nonumber \\
&\leq &\left| \int_{0}^{s}du\Vert [ \alpha _{-u}\left( \left. 
\frac{d\alpha_{-t}(N_{[-L,0]})}{dt}\right| _{t=0}\right) ,J_{-M}]\Vert
\right| 
 \nonumber \\
&\leq &
\left| \int_{0}^{s}du\Vert [ \alpha _{-u}(j_{0,1}).J_{-M}]\Vert \right|
+\left| 
\int_{0}^{s}du\Vert [ \alpha _{-u}(j_{-L-1,-L}),J_{-M}]\Vert \right|.
\label{eq4}
\end{eqnarray}
The last line of (\ref{eq4}) is thanks to (\ref{cont}). 
By translating $J_{-M}$
to neighborhood of the origin, $J_{0}:=\tau _{M}(J_{-M})\in {\cal A}([-1,1])$%
, we can use lemma \ref{th:gp} to estimate the first term of (\ref{eq4}) as 
\begin{eqnarray*}
&&\Vert [ \alpha _{-u}(j_{0,1}),J_{-M}]\Vert 
=\Vert [ \tau_{M}\circ \alpha _{-u}(j_{0,1}),J_{0}]\Vert  \\
&\leq &2\Vert j_{0,1}\Vert \Vert J_{-M}\Vert (N+1)^{5}6\mbox{exp}\left\{-|u|
\left(\frac{M-5}{|u|}-2V(h_{0,1})\right)\right\}
\end{eqnarray*}
In the same manner we obtain the bound for second term of (\ref{eq4}) as 
\begin{eqnarray*}
&&\Vert [ \alpha _{-u}(j_{-L-1,-L}),J_{-M}]\Vert =\Vert [
j_{-L-1,-L},\alpha _{u}\circ \tau _{L-M}(J_{0})]\Vert  \\
&\leq &2\Vert j_{0,1}\Vert \Vert J_{0}\Vert (N+1)^{5}6
\mbox{exp}\left\{-|u|\left(\frac{L-M-5%
}{|u|}-2V(h_{0,1})\right)\right\}.
\end{eqnarray*}
Combination of the above estimates leads 
\begin{eqnarray}
(\ref{eq4})\leq \frac{e^{2V(h_{0,1})|s|}-1}{2V(h_{0,1})}2\Vert j_{0,1}\Vert
\Vert J_{0}\Vert (N+1)^{5}6(e^{-M}+e^{-(L-M)})e^{5}.
\label{eq22}
\end{eqnarray}
Therefore, from (\ref{eq21}) and (\ref{eq22}), we obtain 
\begin{eqnarray}
&&\Vert [ N_{[-L,0]},H_{[-M,M+1]}(t)]
-[N_{[-L,0]},H_{[-M,M+1]}]\Vert  
\nonumber \\
&\leq
&\left| \int_{0}^{t}ds(\Vert [ N_{[-L,0]},J_{M+1}(s)]\Vert +\Vert [
N_{[-L,0]},J_{-M}(s)]\Vert )\right|  
\leq Z_{M,L}(t),  \label{zml}
\end{eqnarray}
where 
\begin{eqnarray*}
Z_{M,L}(t):= &&6(N+1)^{4}\Vert n_{0}\Vert \Vert J_{0}\Vert \frac{e^{-M}}{%
1-e^{-1}}e^{3}\frac{e^{2V(h_{0,1})|t|}-1}{2V(h_{0,1})} \\
&&+2\Vert j_{0,1}\Vert \Vert J_{0}\Vert (N+1)^{5}6e^{5} \\
&&\left(
e^{-M}+e^{-(L-M)}\right)
\frac{1}{2V(h_{0,1})}\left(\frac{e^{2V(h_{0,1})|t|}-1}{%
2V(h_{0,1})}-|t|\right).
\end{eqnarray*}
The integration of (\ref{zml}) with the function $f$ derives 
\begin{eqnarray*}
&&\Vert \int
dti[N_{[-L,0]},H_{[-M,M+1]}(t)]f(t)
-i[N_{[-L,0]},H_{[-M,M+1]}]f(t)\Vert  \\
&\leq &\int dt\Vert [
N_{[-L,0]},H_{[-M,M+1]}(t)]-[N_{[-L,0]},H_{[-M,M+1]}]\Vert |f(t)| \\
&=&\int_{-T}^{T}dt\Vert [
N_{[-L,0]},H_{[-M,M+1]}(t)]
-[N_{[-L,0]},H_{[-M,M+1]}]\Vert |f(t)| \\
&\leq &\left(\int dt|f(t)|^{2}\right)^{1/2}\left(\int_{-T}^{T}dt\Vert [
N_{[-L,0]},H_{[-M,M+1]}(t)]-[N_{[-L,0]},H_{[-M,M+1]}]\Vert ^{2}
\right)^{1/2} \\
&\leq &\left(\int dt|f(t)|^{2}\right)^{1/2}\left(
2\int_{0}^{T}dtZ_{M,L}(t)^{2}\right)^{1/2} \\
&\leq &\left(\int
dt|f(t)|^{2}\right)^{1/2}%
\left\{
A(T)e^{-2M}+B(T)(e^{-2M}+e^{-2(L-M)}+2e^{-L})+C(T)(e^{-2M}+e^{-L})
\right\}^{1/2}
\end{eqnarray*}
where $A(T),B(T)$ and $C(T)$ do not depend upon $M$ and $L$. Consequently we
obtain the following: 
\begin{eqnarray*}
\lim_{M\to \infty }\lim_{L\to \infty }\int dti[\hat{N}%
_{[-L,0]},\hat{H}_{[-M,M+1]}(t)]f(t)=\sqrt{2\pi }j_{0,1}\tilde{f}(0).
\end{eqnarray*}
Thus the proof is completed. \hfill {\bf Q.E.D.} 

This lemma gives a starting point for our discussion. Note that
the ordering of limiting procedures, $L \to \infty$ and $M\to \infty$,
cannot be changed.
 In fact one
can easily see that if one takes $M\to \infty$ first, the left hand side of
the above lemma vanishes. 

To study the property of energy momentum spectrum, a proper
correlation function should be investigated. 

\begin{definition}
To investigate the property of $E_{\omega}(dkd\epsilon)$ we
define a ``function" $\tilde{\rho}(k,\epsilon)$ as 
\begin{eqnarray*}
\tilde{\rho}(k,\epsilon)dk d\epsilon =(\Omega,i\hat{n}_0 E_{\omega}(dk
d\epsilon)\hat{h}_{0,1}\Omega).
\end{eqnarray*}
Precisely $\tilde{\rho}$ is a distribution over 
infinitely differentiable function of $k$ and $\epsilon$.
To get rid of 
an effect of a product of expectations $\omega(n_0)\omega(h_{0,1})$, 
we again use the 
notation $\hat{A}:=A-\omega(A)$.
\end{definition}
The following is the main theorem. 
\begin{theorem}\label{th:main} 
For $\omega$, a state with nonvanishing current, the energy spectrum has
singularity at the origin. i.e., 
\begin{eqnarray*}
- 2\pi i \left. \left(
\frac{\partial}{\partial k}\tilde{\rho}(k,\epsilon)+  \frac{%
\partial}{\partial k}\tilde{\rho}(-k,-\epsilon)^*\right)\right|_{k=0}
=\omega(j_{0,1})\delta(\epsilon)
\end{eqnarray*}
holds. 
\end{theorem}

{\bf Proof}\newline
Since what we are interested in is the spectrum property with respect to $%
\omega $, we take an expectation value for $\omega $ of the above lemma \ref
{th:start}. 
\begin{eqnarray}
\lim_{M\to \infty }\lim_{L\to \infty }\int dt(\Omega ,i[\hat{%
N}_{[-L,0]},\hat{H}_{[-M,M+1]}(t)]f(t)\Omega )=\sqrt{2\pi }\omega (j_{0,1})%
\tilde{f}(0),
\end{eqnarray}
The information with respect to the energy-momentum spectrum is encoded in
the left hand side of the above equation. To draw it we define functions $%
r_{L}$ and $s_{M}$ as 
\begin{eqnarray*}
r_{L}(x)&:=&1\ \mbox{for}\ -L\leq x\leq 0,\ \mbox{otherwise}\ 0 \\
s_{M}(x)&:=&1\ \mbox{for}\ -M\leq x\leq M,\ \mbox{otherwise}\ 0.
\end{eqnarray*}
These objects are used to derive 
\begin{eqnarray}
\int dt(\Omega ,i[\hat{N}_{[-L,0]},\hat{H}_{[-M,M+1]}(t)]\Omega )f(t)
\nonumber \\
=\int
dt\sum_{x}\sum_{y}r_{L}(x)s_{M}(y)(\Omega ,i[\hat{n}_{x}(0),\hat{h}%
_{y,y+1}(t)]\Omega )f(t).
\label{eq5}
\end{eqnarray}
By use of the spectrum decomposition of the space-time translation unitary
operator $U(z,t):=\int e^{i(\epsilon t-kz)}E_{\omega }(dkd\epsilon )$, we
denote Fourier transform of $\tilde{\rho}(k,\epsilon )$ as 
\begin{eqnarray}
\rho (z,t):=\frac{1}{2\pi \sqrt{2\pi }}\int d\epsilon \int_{-\pi }^{\pi }dk%
\tilde{\rho}(k,\epsilon )e^{i(kz-\epsilon t)}=\frac{1}{2\pi \sqrt{2\pi }}%
(\Omega ,i\hat{n}_0 \hat{h}_{-z,-z+1}(-t)\Omega ),
\end{eqnarray}
then we can write the equation (\ref{eq5}) as 
\begin{eqnarray}
(\ref{eq5})=4\pi \sqrt{2\pi }\int dt\sum_{z}Re\left(\rho
(z,-t)(\sum_{x}r_{L}(x)s_{M}(x-z))\right)f(t)
\end{eqnarray}
Now the relation 
\begin{eqnarray}
\lim_{L\to \infty}\sum_{x}r_{L}(x)s_{M}(x-z)=\left\{ 
\begin{array}{rl}
2M+1, & \quad z<-M \\ 
M+1-z, & \quad -M\leq z\leq M \\ 
0, & \quad M<z
\end{array}
\right. 
\end{eqnarray}
is used to show the limiting value for $L$ to infinity as 
\begin{eqnarray}
\lim_{L\to \infty } &&\int dt(\Omega ,i[\hat{N}_{[-L,0]},\hat{H}%
_{[-M,M+1]}(t)]\Omega )f(t)  \nonumber \\
&=&4\pi \sqrt{2\pi }\int dtf(t)Re((\sum_{z<-M}\rho (z,-t)(2M+1))  \label{eq8}
\\
&+&4\pi \sqrt{2\pi }\int dtf(t)Re((\sum_{-M\leq z\leq M}\rho (z,-t)(M+1))
\label{eq9} \\
&+&4\pi \sqrt{2\pi }\int dtf(t)Re((-\sum_{-M\leq z\leq M}z\rho (z,-t)).
\label{eq10}
\end{eqnarray}
Next consider what will occur when $M$ is made infinity in the above
equation. In the following, we show that (\ref{eq8}) and (\ref{eq9})
approach zero as $M\to \infty $. Let us begin with (\ref{eq8}), 
\begin{eqnarray*}
(\ref{eq8}) &=&2\int dtf(t)Re(\sum_{z>M}(\Omega ,i\hat{n}_{0}\hat{h}%
_{z,z+1}(t)\Omega ))(2M+1)) \\
&=&
i\int dtf(t)(\Omega ,[\hat{n}_{0},\sum_{z>M}\hat{h}_{z,z+1}(t)]\Omega )(2M+1).
\end{eqnarray*}
Thanks to Cauchy-Schwarz inequality, one can obtain 
\begin{eqnarray*}
|(\ref{eq8})| &\leq &\int dt|f(t)||(\Omega ,[\hat{n}_{0},\sum_{z>M}\hat{h}%
_{z,z+1}(t)]\Omega )|(2M+1) \\
&\leq &(2M+1)\left(\int dt|f(t)|^{2}\right)^{1/2}
\left(\int_{-T}^{T}dt\Vert [ \hat{n}%
_{0},\sum_{z>M}\hat{h}_{z,z+1}(t)]\Vert ^{2}\right)^{1/2}.
\end{eqnarray*}
Since as for the integrand of the above equation, the group-velocity lemma 
\ref{th:gp} is used to show 
\begin{eqnarray}
\Vert [ \hat{n}_{0},\hat{h}_{z,z+1}(t)]\Vert \leq 2(N+1)^{3}\Vert \hat{n}%
_{0}\Vert \Vert \hat{h}_{0,1}\Vert 2\mbox{exp}\left\{
-|t|\left(\frac{|z|-3}{|t|}%
-2V(h_{0,1})\right)\right\}
\end{eqnarray}
and 
\begin{eqnarray}
\Vert [ \hat{n}_{0},\sum_{z>M}\hat{h}_{z,z+1}(t)]\Vert \leq
2(N+1)^{3}\Vert \hat{n}_{0}\Vert \Vert \hat{h}_{0,1}\Vert
2e^{3}e^{2|t|V(h_{0,1})}\frac{e^{-M}}{e-1}
\end{eqnarray}
Thus finally we obtain 
\begin{eqnarray}
|(\ref{eq8})|\leq \left(\int dt|f(t)|^{2}\right)^{1/2}
(2M+1)2(N+1)^{3}\Vert \hat{n}%
_{0}\Vert \Vert \hat{h}_{0,1}\Vert 2e^{3}\frac{e^{-M}}{e-1}\left(\frac{%
e^{4V(h_{0,1})T}-1}{2V(h_{0,1})}\right)^{1/2}
\end{eqnarray}
which approaches zero as $M\to \infty $.

 Next we estimate the equation(\ref{eq9}), 
\begin{eqnarray*}
|(\ref{eq9})| \leq (M+1) \left(\int dt |f(t)|^2\right)^{1/2} 
\left(\int^{T}_{-T} dt |(\Omega,[%
\hat{n}_0,\sum_{z=-M}^M \hat{h}_{z,z+1}(t)] \Omega)|^2 \right)^{1/2}.
\end{eqnarray*}
The integrand of the above equation can be written by use of stationarity of 
$\omega$ as 
\begin{eqnarray*}
(\Omega,[\hat{n}_0,\sum_{z=-M}^M 
\hat{h}_{z,z+1}(t)]\Omega) &=&(\Omega,[\hat{n}%
_0,\sum_{-M}^M \hat{h}_{z,z+1}(0)]\Omega) +\int^t_0 ds 
(\Omega,[\hat{n}_0,\frac{d}{%
ds}\hat{H}_{[-M,M+1]}(s)]\Omega)  \nonumber \\
&=& \int^t_0 ds (\Omega,[\hat{n},\alpha_s\left(
\left. \frac{d\hat{H}_{-M,M+1]}(t)}{%
dt}\right|_{ t=0}\right)]\Omega)  \nonumber \\
&=& \int^t_0 ds (\Omega,[\hat{n},J_{-M}(s)-J_{M+1}(s)]\Omega).
\end{eqnarray*}
As before, decomposition into energy current terms 
\begin{eqnarray*}
\frac{d\hat{H}_{[-M,M+1]}}{dt}
=-[H_{[-M,M+1]},H_{[-M-1,M+2]}] =J_{-M}-J_{M+1},
\end{eqnarray*}
where $J_{-M}\in{\cal A} ([-M-1,-M+1])$ and 
$J_{M+1}\in {\cal A} ([M,M+2])$
leads 
\begin{eqnarray*}
|(\Omega,[\hat{n}_0,\sum_{z=-M}^M \hat{h}_{z,z+1}(t)]\Omega)| 
\leq \left|\int^t_0 ds \Vert
[\hat{n}_0,J_{-M}(s)]\Vert \right| +
\left| \int^t_0 ds \Vert [\hat{n}_0, J_{M+1}(s)]\Vert\right|.
\end{eqnarray*}
And the following estimations which are obtained by direct use of
group-velocity lemma \ref{th:gp} 
\begin{eqnarray*}
\Vert [\hat{n}_0,J_{-M}(s)]\Vert &\leq& 2(N+1)^{4}\Vert n_0\Vert \Vert
J_{0}\Vert 3 \mbox{exp}\left\{\left(-|s|(\frac{M-4}{|s|}-2V(h_{0,1})\right)
\right\}  \nonumber \\
\Vert [\hat{n}_0,J_{M+1}(s)]\Vert &\leq& 2(N+1)^{4}\Vert n_0\Vert \Vert
J_{0}\Vert 3 \mbox{exp}\left\{\left(-|s|(\frac{M-4}{|s|}-2V(h_{0,1})
\right)\right\}  \nonumber
\end{eqnarray*}
thus it leads, 
\begin{eqnarray*}
|(\Omega,[\hat{n}_0,\sum_{z=-M}^M \hat{h}_{z,z+1}(t)]\Omega)| 
\leq 2(N+1)^{4} \Vert
n_0 \Vert \Vert J_0 \Vert 3 e^{4} e^{-M} \frac{e^{2V(h_{0,1})|t|}-1}{%
V(h_{0,1})}.
\end{eqnarray*}
Finally we obtain 
\begin{eqnarray*}
&&\left(\int^{T}_{-T} dt|(\Omega,[\hat{n}_0,\sum_{z=-M}^M \hat{h}%
_{z,z+1}(t)
\Omega)|^2\right)^{1/2} \leq 2(N+1)^{4} \Vert n_0 \Vert \Vert J_0 \Vert 3
e^{4} e^{-M}  \nonumber \\
&& \times \frac{1}{V(h_{0,1})} \sqrt{\frac{1}{2V(h_{0,1})}}
(e^{4V(h_{0,1})T}-4e^{2V(h_{0,1})T}+4TV(h_{0,1}))^{1/2}
\end{eqnarray*}
and can see 
\begin{eqnarray*}
\lim_{M \to \infty}(\ref{eq9}) =0
\end{eqnarray*}
holds. Now we estimate the equation(\ref{eq10}) as 
\begin{eqnarray}
(\ref{eq10})&=& 2\pi \sqrt{2\pi} \int dt f(t) (-\sum_{-M\leq z \leq
M}z\rho(z,-t) -\sum_{-M\leq z \leq M} z \rho(z,-t)^*)  \nonumber \\
&=& -i \int d\epsilon \int dt \int dk f(t) \sum_{-M \leq z \leq
M}e^{i(\epsilon t+kz)} \frac{\partial}{\partial k} (\tilde{\rho}(k,\epsilon)+%
\tilde{\rho}(-k,-\epsilon)).
\end{eqnarray}
An equation, $\lim_{M\to \infty} \sum_{-M\leq z\leq M} e^{ikz}
=2 \pi \delta(k)$, 
derives the following relation for $M \to \infty$,
\begin{eqnarray}
\lim_{M\to\infty}(\ref{eq10}) =-i \sqrt{2 \pi}2 \pi\int d\epsilon \tilde{f}%
(\epsilon) \left.\left(\frac{\partial}{\partial k}\tilde{\rho}(k,\epsilon)+ 
\frac{\partial}{\partial k}\tilde{\rho}(-k,-\epsilon)^*\right)\right|_{k=0}
\end{eqnarray}
Finally we obtain the following equation, 
\begin{eqnarray}
-2 \pi i \left. 
\left(\frac{\partial}{\partial k}\tilde{\rho}(k,\epsilon)+ \frac{%
\partial}{\partial k}\tilde{\rho}(-k,-\epsilon)^*\right)\right|_{k=0}
=\omega(j_{0,1})\delta(\epsilon).
\end{eqnarray}
The proof is completed. \hfill {\bf Q.E.D.} 
\section{ Conclusion and Outlook}

We considered states over one-dimensional infinite lattice
which are stationary, translationally invariant and have non-vanishing
current expectations. The spectrum of space-time translation unitary
operator with respect to such a state was investigated and was shown to have
singularity at the origin $(k,\epsilon)=(0,0)$. 
The theorem is a consequence of only the nonvanishingness of current
expectation, and we do not know whether physically more natural conditions
give more detail information of the spectrum. 
It is also interesting to investigate whether our 
result can be generalized to higher dimensional lattices. 
{\bf Acknowledgment} \newline
I would like to thank Izumi Ojima, Yoshiko Ogata and anonymous 
referees for helpful
discussions and comments. 

\end{document}